\documentclass[journal]{IEEEtran}
\usepackage{changes}
\usepackage{amsmath,amsfonts}
\usepackage{algorithmic}
\usepackage{algorithm}
\usepackage{array}
\usepackage[caption=false,font=footnotesize,labelfont=sf,textfont=sf]{subfig}
\usepackage{textcomp}
\usepackage{enumerate}
\usepackage{stfloats}
\usepackage{url}
\usepackage{verbatim}
\usepackage{graphicx}
\usepackage{cite}
\usepackage{color}
\usepackage{multirow}
\usepackage{tikz}
\usetikzlibrary{positioning, shapes.geometric}
\usepackage{adjustbox}
\usepackage[flushleft]{threeparttable}
 \usepackage{pifont}

\begin{document}

\title{Study of Class-Incremental Radio Frequency Fingerprint Recognition Without Storing Exemplars}

\author{Rundong Jiang, Jun Hu, Yunqi Song, Zhiyuan Xie, Shiyou Xu
	
	\thanks{All authors are with the School of Electronics and Communication Engineering, Sun Yat-Sen University, Shenzhen 518000, China.
 }}

\markboth{JOURNAL,~Vol.~14, No.~8, JANUARY~2023}%
{Shell \MakeLowercase{\textit{et al.}}: A Sample Article Using IEEEtran.cls for IEEE Journals}


\maketitle

\begin{abstract}
    The rapid proliferation of wireless devices makes robust identity authentication essential. Radio Frequency Fingerprinting (RFF) exploits device-specific, hard-to-forge physical-layer impairments for identification, and is promising for IoT and unmanned systems. In practice, however, new devices continuously join deployed systems while per-class training data are limited. Conventional static training and naive replay of stored exemplars are impractical due to growing class cardinality, storage cost, and privacy concerns.
    We propose an exemplar-free class-incremental learning framework tailored to RFF recognition. Starting from a pretrained feature extractor, we freeze the backbone during incremental stages and train only a classifier together with lightweight Adapter modules that perform small task-specific feature adjustments. For each class we fit a diagonal Gaussian Mixture Model (GMM) to the backbone features and sample pseudo-features from these fitted distributions to rehearse past classes without storing raw signals. To improve robustness under few-shot conditions we introduce a time-domain random-masking augmentation and adopt a multi-teacher distillation scheme to compress stage-wise Adapters into a single inference Adapter, trading off accuracy and runtime efficiency.
    We evaluate the method on large, self-collected ADS-B datasets: the backbone is pretrained on 2,175 classes and incremental experiments are run on a disjoint set of 669 classes with multiple rounds and step sizes. Against several representative baselines (both exemplar-based and exemplar-free), our approach consistently yields higher average accuracy and lower forgetting, while using substantially less storage and avoiding raw-data retention.
    The proposed pipeline is reproducible and provides a practical, low-storage solution for RFF deployment in resource- and privacy-constrained environments.

\end{abstract}

$\mathbf{Keywords:}$ Class-incremental learning, RF fingerprinting, Feature fitting

This work has been submitted to the IEEE for possible publication. Copyright may be transferred without notice, after which this version may no longer be accessible.

\section{Introduction}
Radio frequency fingerprints (RF fingerprints) are unique characteristics produced by hardware variations during signal transmission, such as frequency offsets, modulation errors, and transient responses. These subtle differences arise from manufacturing variations of components and can serve as device "identity" markers. Through signal analysis and machine learning, RF fingerprinting enables device identification, authentication, and anti-cloning tracing, effectively combating cloned devices and network fraud; it has important applications in IoT security and spectrum management.

Existing RFF recognition methods mostly focus on feature extraction. For neural-network-based approaches, models are typically trained once on a training set and then deployed for inference. In practical scenarios, devices continuously join the network, so the number of classes the model must recognize increases over time. In such settings, re-deploying or re-training at the edge often requires storing all old-device samples, which introduces large storage costs and privacy risks.

Therefore, enabling incremental learning for RFF—i.e., continually learning new device classes without retaining old device samples—is an important problem.

Compared to other classification tasks, class-incremental learning for RFF has characteristics such as a large number of devices and classes and small inter-device differences. In identity authentication scenarios, RFF needs to distinguish many devices; because RFF focuses on extracting features from signals with the same frequency and modulation, it is a fine-grained recognition problem \cite{shenExplorationTransferableDeep2024,zhangFineGrainedRadioFrequency2023,zhaoRadioFrequencyFingerprinting2023}. Thus, using a pretrained model as a feature extractor followed by fine-tuning is a reasonable choice; the problem reduces to incremental training of the classifier. Methods that directly store some original samples face storage and privacy limitations in practice.

When a pretrained stage has covered sufficiently many devices and channel variations, and new incremental classes do not exhibit significant domain shift from the pretraining distribution, the pretrained feature extractor can be treated as a relatively stable representation. Under this fine-grained premise, the pretrained model can extract RFF features; by fixing the pretrained parameters and applying lightweight Adapters for incremental fine-tuning while training the classifier incrementally, the key challenge becomes classifier incremental training.

Classifier incremental training focuses on preserving decision boundaries. From the perspective of decision boundaries and data preservation, we propose fitting feature distributions and regenerating pseudo-features during incremental training. After each stage, we extract features for the classes present in that stage, fit distributions per class, and sample pseudo-features from those distributions during later training to rehearse past classes. We use a diagonal Gaussian Mixture Model to fit feature vectors, keeping only diagonal elements of covariance matrices to simplify computation and storage; this sacrifices some fitting fidelity for greater stability.

For each incremental task, we also add a fine-tuning layer called the Adapter to adjust feature distribution shifts introduced by new classes. During training, we only use current-task class samples to train the classifier and the Adapter. A separate Adapter is trained for each task; at inference time, since the stage identity of an unknown sample is not available a priori, we distill all historical Adapters into a single Adapter to avoid requiring prior knowledge of a sample's task.

We adopt the twin-structure feature-difference network-based feature extractor plus classifier architecture proposed in \cite{jiangGeneralizedRadioFrequency2024} as the pretrained model. Extensive experiments are conducted on two self-collected large ADS-B datasets: the pretrained model is trained with 2,175 classes, and multi-step incremental training is performed on an additional 669 classes that are disjoint from the pretraining set. During training, we apply a random-masking augmentation that masks random lengths at the start or end of time-domain signals, treating the masked result as an additional same-class sample; this increases feature diversity and model robustness.

Experiments compare our method with mainstream baselines, showing advantages in storage usage and superiority in accuracy and forgetting, demonstrating stronger continual learning capability. Our method leverages existing model parameters and knowledge to avoid storing original exemplar samples, making it friendly to privacy-sensitive scenarios.

The main contributions are summarized as follows:
\begin{itemize}
    \item We present an exemplar-free class-incremental learning framework for RFF recognition that preserves prior-class decision boundaries by sampling pseudo-features from compact per-class models, eliminating the need to retain raw exemplars.
    \item We adopt diagonal-covariance GMMs to parameterize per-class backbone features in high-dimensional, few-shot regimes; this reduces covariance storage from $O(D^2)$ to $O(D)$ per component and improves fitting stability.
    \item We introduce lightweight Adapter modules for stage-wise adaptation and a multi-teacher distillation procedure that merges historical Adapters into a single student Adapter for efficient inference.
    \item We conduct extensive experiments on large, self-collected ADS-B datasets (2,175 classes for pretraining; 669 disjoint classes for incremental evaluation) and show that the proposed method outperforms representative baselines in average accuracy and forgetting while requiring far less storage.
\end{itemize}

\section{Related Work}
Class-incremental learning (CIL) is a typical scenario in continual learning where the model sequentially learns tasks or classes without accessing original data from previous tasks, gradually expanding its recognition capability. The main challenge is catastrophic forgetting, where learning new classes degrades performance on prior classes. Existing work can be grouped into three categories: regularization-based methods, replay-based methods, and distillation/feature-preservation methods.

\subsection{Regularization-based Methods}
Regularization methods constrain parameter updates to preserve knowledge of previous tasks. Elastic Weight Consolidation (EWC) estimates parameter importance using the Fisher information matrix and penalizes changes to critical parameters \cite{aichElasticWeightConsolidation2021}. Other approaches limit updates along low-forgetting directions to maintain adaptability for new tasks \cite{wenClassIncrementalLearning2024}. However, these methods often rely on task boundary information and suffer from inaccurate Fisher estimates. As tasks accumulate, parameter plasticity diminishes, reducing adaptability to new tasks.

\subsection{Replay and Prototype Memory}
Replay methods retain subsets of old-class samples or compressed representations for rehearsal. iCaRL, a representative approach, maintains class exemplars and combines a nearest-mean classifier with feature-space replay \cite{rebuffiICaRLIncrementalClassifier2017}. Subsequent methods, such as End-to-End Incremental Learning \cite{castroEndtoEndIncrementalLearning2018} and Dark Experience Replay \cite{buzzegaDarkExperienceGeneral2020}, refine loss functions and memory update strategies. Some approaches dynamically expand representational capacity to improve old-class retention \cite{yanDynamicallyExpandableRepresentation2021}.

Generative replay methods synthesize old-class samples using GANs or VAEs \cite{wuMemoryReplayGANs2019,vandevenClassIncrementalLearningGenerative2021,chaiMalFSCILFewShotClassIncremental2025}. Other techniques reconstruct pseudo-features through linear combinations of known features \cite{zhangForwardBackwardCompatible2024}. However, generative models increase storage and training complexity and are challenging to train under few-shot conditions. Linear combination methods assume convexity in feature space, which may not generalize well.

\subsection{Knowledge Distillation and Feature Preservation}
Knowledge distillation transfers knowledge from the old model to the new model, retaining performance on previous classes. LwF first applied distillation to incremental learning \cite{liuMnemonicsTrainingMultiClass2020}. Prototype-aware contrastive distillation methods constrain output similarity and use prototype-based contrastive loss to improve feature compatibility \cite{yangProCoPrototypeawareContrastive2022}. Dark Experience Replay stores intermediate features for implicit knowledge reuse. Methods like LUCIR add cosine-based constraints to enhance discrimination between old and new classes \cite{houLearningUnifiedClassifier2019}.

The use of large pretrained models in class-incremental learning has gained attention \cite{zhouExpandableSubspaceEnsemble2024,rothPractitionersGuideContinual2024,zhouDualConsolidationPreTrained2024}. DyTox employs a transformer backbone with dynamic adaptation for task-agnostic continual learning \cite{douillardDyToxTransformersContinual2022}.

\section{Method}
RF fingerprinting aims to distinguish transmitting devices under identical modulation, protocol, and communication conditions. Prior work demonstrates that even when devices use the same protocols and modulation formats, unique RF chain characteristics introduce reproducible differences in emitted signals, enabling device-level discrimination \cite{merchantDeepLearningRF2018}. This makes RFF an inherently fine-grained recognition task, motivating the use of a pretrained feature extractor and classifier adaptation for incremental learning.

Using a high-capacity pretrained extractor, the incremental problem reduces to classifier adaptation \cite{jiangTransmitterIdentificationVolterra2025}. The goal is to preserve the classifier's decision boundaries for old classes while learning new classes. For a $K$-class problem, the posterior is defined as:
\begin{equation}
    P(y=c\mid\mathbf{x};\theta)
    \label{eq:posterior}
\end{equation}
where $c$ denotes class $c$, $\mathbf{x}\in\mathbb{R}^d$ is the input, $y\in\{1,\dots,K\}$ the label, and $\theta$ the model parameters. The decision boundary between classes $i$ and $j$ is:
\begin{equation}
    B_{i,j}=\{\mathbf{x}\mid P(y=i\mid\mathbf{x};\theta)=P(y=j\mid\mathbf{x};\theta)\}
    \label{eq:decision_boundary}
\end{equation}

To explicitly express decision-boundary preservation across incremental stages, we require:
\begin{equation}
    \mathbf{x}\in B_{i,j}^{(t)}\Rightarrow\mathbf{x}\in B_{i,j}^{(t+1)}
    \label{eq:keep_decision_boundary}
\end{equation}

The incremental objective is to ensure:
\begin{equation}
    P(y=c\mid x;\theta_t)\approx P(y=c\mid x;\theta_{t-1})
\end{equation}
so that decision boundaries are preserved. Using KL divergence between old and new posteriors yields a distillation loss:
\begin{equation}
    \mathcal{L}_{\mathrm{distil}}=\operatorname{KL}\left(P_{t}(y\mid\mathbf{x};\theta_t)\mid\mid P_{t+1}(y\mid\mathbf{x};\theta_{t+1})\right)
    \label{eq:kl_div}
\end{equation}

\begin{figure}
    \centering
    \includegraphics[width=8.5cm]{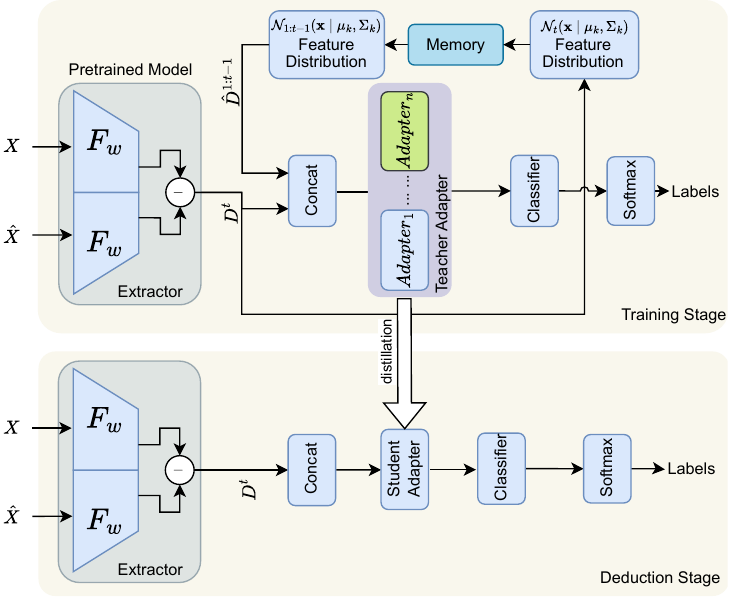}
    \caption{Overall framework}
    \label{fig:gmm_diagram}
\end{figure}

We build on the twin-structured feature-difference extractor plus classifier architecture in \cite{jiangGeneralizedRadioFrequency2024}, and propose an exemplar-free class-incremental method illustrated in Fig.~\ref{fig:gmm_diagram}.
The pretrained extractor computes:
\begin{equation}\label{eq:diff_feature}
    D^t=F_w(X)-F_w(\hat{X})
\end{equation}
where $F_w(\cdot)$ is the extractor, $X$ the input sample, $\hat{X}$ a paired reconstruction without hardware fingerprints, and $D^t$ the feature for the $t$-th incremental round.

\begin{figure}[b]
    \centering
    \includegraphics[width=4.5cm]{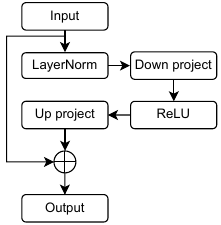}
    \caption{Structure of Adapter}
    \label{fig:adapter_structure}
\end{figure}

The incremental process includes training and inference stages. After each training stage, we fit a diagonal GMM to the features of each class, obtaining per-class feature distributions $\mathcal{N}_t$. During later incremental training, we sample pseudo-features $\hat{D}_{t-1}$ from these fitted distributions and train the classifier with both new-class features and sampled pseudo-features to preserve old decision boundaries. Each round also adds an Adapter module to adjust for feature drift \cite{pfeifferAdapterFusionNonDestructiveTask2021}; the extractor features pass through the Adapter before classification. The Adapter structure is shown in Fig.~\ref{fig:adapter_structure}.

\begin{figure}[b]
    \centering
    \includegraphics[width=3cm]{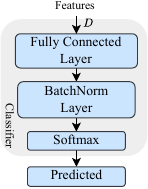}
    \caption{Classifier structure}
    \label{fig:classifier_structure}
\end{figure}

During inference, the sample's task is unknown, so all historical Adapters are fused via distillation: historical Adapters act as teachers, and a new Adapter is trained as the student by minimizing output differences. The fitted pseudo-features enable rehearsal without storing raw exemplars. We adopt a fully-connected classifier with batch normalization as shown in Fig.~\ref{fig:classifier_structure}.

While nearest-mean classifiers (NMC) keep a mean vector per class and classify by closest center \cite{snellPrototypicalNetworksFewShot2017}, they implicitly assume roughly spherical feature clusters. In non-ideal feature geometries, center-based decisions can shift boundaries and hurt open-set detection, which is critical in RFF scenarios.

\begin{figure*}[h]
    \centering
    \subfloat[Real samples feature visualization\label{subfig:tsne_orig}]{\includegraphics[width=5.7cm]{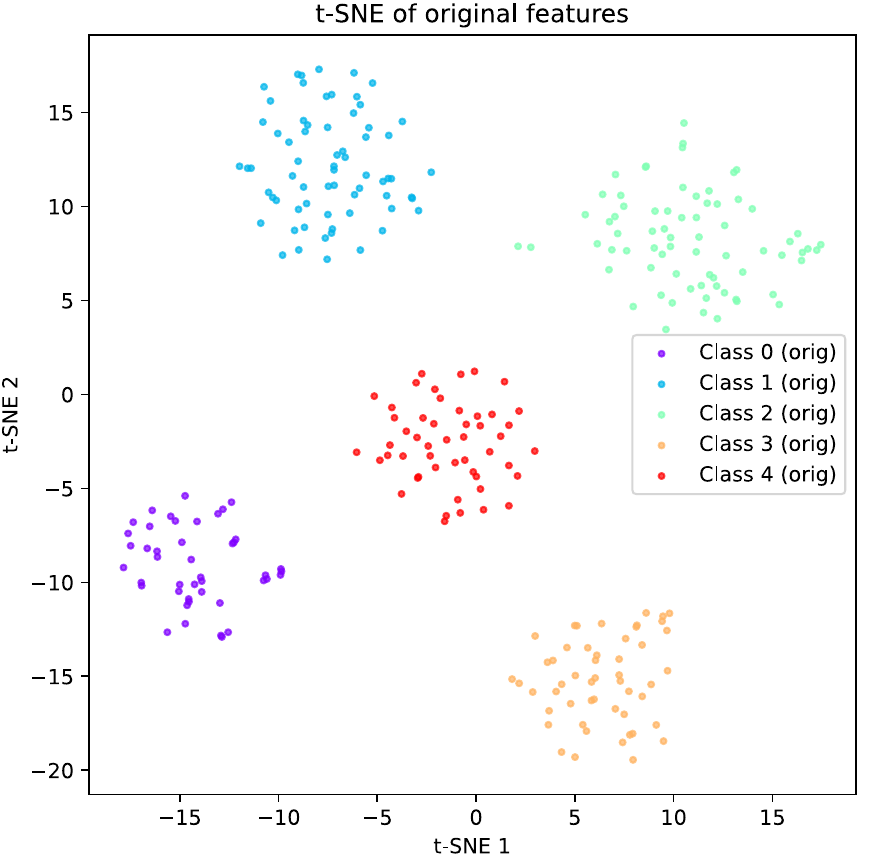}}
    \subfloat[Visualization of pseudo-features sampled from the GMM\label{subfig:tsne_pseudo}]{\includegraphics[width=5.7cm]{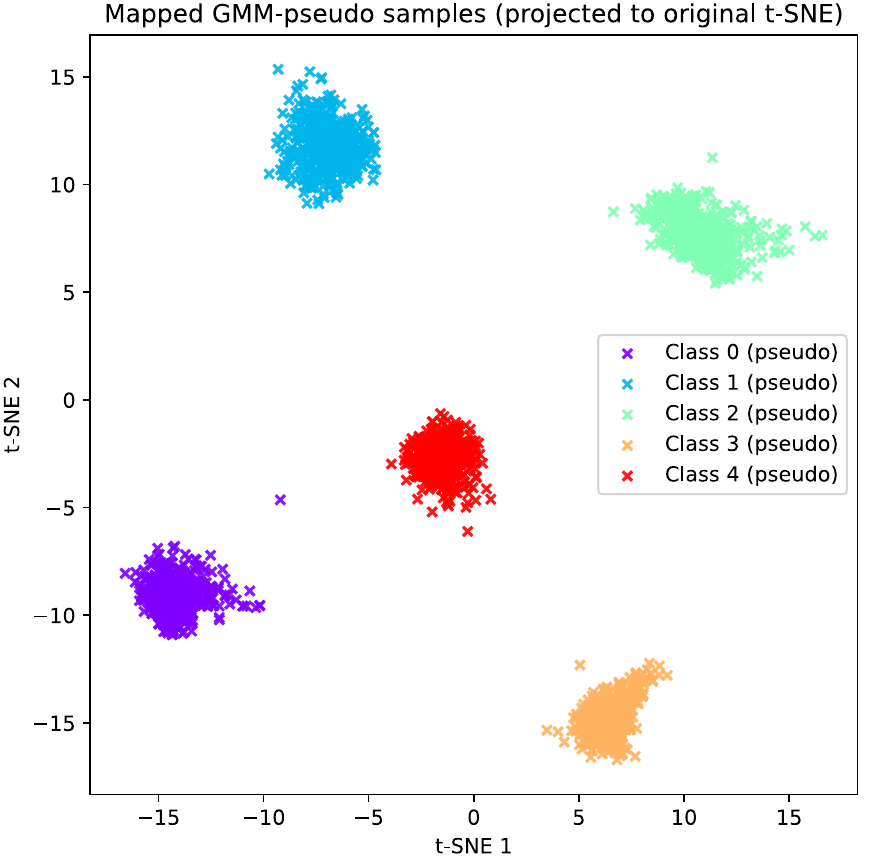}}
    
    \subfloat[Overlay of both visualizations\label{subfig:tsne_overlay}]{\includegraphics[width=6cm]{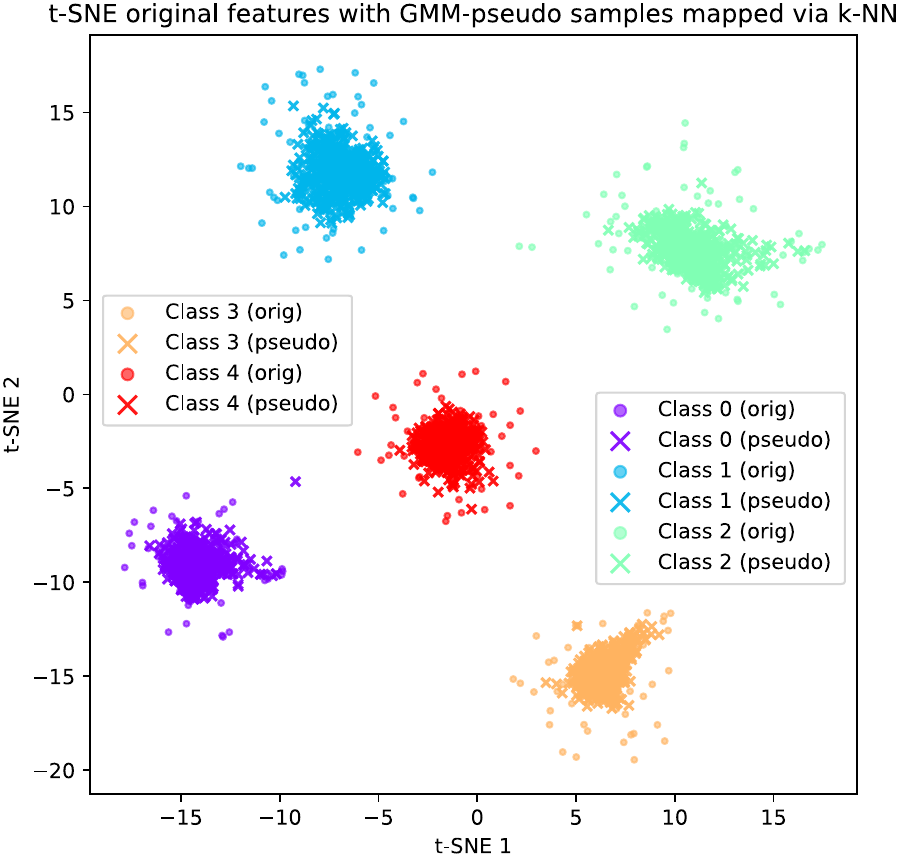}}
    \caption{t-SNE comparison of real and GMM-sampled pseudo-features}
    \label{fig:features_tsne}
\end{figure*}

\subsection{Feature Distribution Fitting}
\begin{figure}[b]
    \centering
    \includegraphics[width=6.5cm]{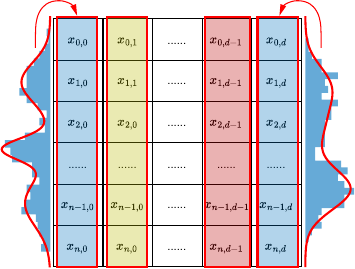}
    \caption{Gaussian Mixture Model}
    \label{fig:gmm_diagram_zhankai}
\end{figure}

We fit the features $D^t$ in Fig.~\ref{fig:gmm_diagram_zhankai} with a multivariate Gaussian mixture model. Directly fitting full-covariance GMMs in high dimensions is computationally expensive: for feature dimension $D$ and $k$ components, the per-component covariance has $\frac{D(D+1)}{2}$ parameters, leading to quadratic scaling. In few-shot settings, this causes poor estimates and overfitting. We therefore use diagonal-covariance GMMs, reducing per-component covariance parameters to $D$ and total parameters to scale linearly with $D$.

The diagonal GMM is defined as:
\begin{equation}
    p(\mathbf{x})=\sum_{k=1}^{K}\pi_{k}\cdot{\mathcal{N}}(\mathbf{x}\mid\mu_{k},\Sigma_{k})
    \label{eq:gmm_}
\end{equation}
with diagonal $\Sigma_k=\mathrm{diag}(\sigma_{k,1}^2,\dots,\sigma_{k,d}^2)$, and:
\begin{equation}
    {\mathcal{N}}(\mathbf{x}\mid\mu_{k},\Sigma_{k})=\prod_{i=1}^{d}\mathcal{N}(x_i\mid\mu_{k,i},\sigma_{k,i}^2)
    \label{eq:gaussian_gmm}
\end{equation}

Equivalently, the density admits the closed form:
\begin{equation}
    \begin{aligned}
    {\mathcal{N}}(\mathbf{x}\mid\mu_{k},\Sigma_{k})=&{\frac{1}{(2\pi)^{d/2}\prod_{i=1}^{d}\sigma_{k,i}}}\cdot \\
    &\exp\left(-{\frac{1}{2}}\sum_{i=1}^{d}{\frac{(x_{i}-\mu_{k,i})^{2}}{\sigma_{k,i}^{2}}}\right)
    \end{aligned}
    \label{eq:gaussian_gmm_zhankai}
\end{equation}

We fit the GMM using the EM algorithm:
\begin{enumerate}
    \item Compute responsibilities $\gamma(z_{nk})=\frac{\pi_k N(\mathbf{x}_n\mid\mu_k,\Sigma_k)}{\sum_j\pi_j N(\mathbf{x}_n\mid\mu_j,\Sigma_j)}$.
    \item Update $\pi_k^{\textrm{new}}=N_k/N$, where $N_k=\sum_n\gamma(z_{n,k})$.
    \item Update $\mu_k^{\textrm{new}}=\frac{1}{N_k}\sum_n\gamma(z_{n,k})x_n$.
    \item Update diagonal covariance: $\sigma_{k,i}^2=\frac{1}{N_k}\sum_n\gamma(z_{n,k})(x_{n,i}-\mu_{k,i})^2$.
    \item Repeat steps 1–4 until convergence.
\end{enumerate}

Sampling proceeds by selecting a component according to mixture weights and drawing independent normal samples per dimension.

\subsection{Feature Alignment}
To merge Adapters across stages, we use distillation. A student Adapter is trained using all historical Adapters as teachers by minimizing output discrepancies. During distillation, we add feature-alignment losses, including pointwise MSE and distribution alignment.
\begin{figure}[b]
    \centering
    \includegraphics[width=8cm]{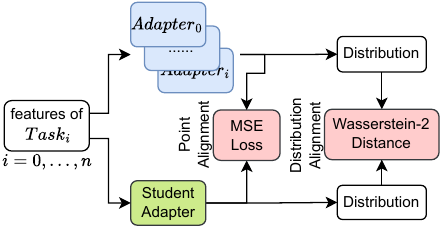}
    \caption{Alignment of features}
    \label{fig:feature_alignment}
\end{figure}
\subsubsection{Pointwise Alignment}
Pointwise alignment uses MSE:
\begin{equation}
    \mathcal{L}_{\mathrm{align}}=\frac{1}{N}\sum_{i=1}^N\|D_i-\hat{D}_i\|_2^2
    \label{eq:feature_alignment}
\end{equation}
which encourages the student Adapter outputs on pseudo-features to match teachers'.

\subsubsection{Distribution Alignment}
For distribution alignment, we use the 2-Wasserstein distance between multivariate Gaussians, which has a closed form for Gaussians:
\begin{equation}
    \begin{aligned}
    &W_2(\mathcal{N}(\mathbf{m}_1,\mathbf{C}_1),\mathcal{N}(\mathbf{m}_2,\mathbf{C}_2))=\\
    &\|\mathbf{m}_1-\mathbf{m}_2\|_2^2+\mathrm{tr}\left(\mathbf{C}_1+\mathbf{C}_2-2(\mathbf{C}_1^{1/2}\mathbf{C}_2\mathbf{C}_1^{1/2})^{1/2}\right)
    \end{aligned}
    \label{eq:wasserstein_alignment}
\end{equation}
which we use to align pseudo-feature distributions during incremental training.

\subsection{Training Data Augmentation}
RFF data are time-domain signals and fingerprint characteristics are independent of payload content. To augment limited samples while minimizing signal distortion, we apply random time-domain masking at the start or end of a sample. Formally:
\begin{equation}
    \tilde{s}[n]=s[n]\cdot g_{n_0}[k]
    \label{eq:mask_from_start_end}
\end{equation}
where $g_{n_0}[k]=u[n-n_0]-u[n-n_0-k]$ is a gate window, $u[n]$ the step function, $n_0$ the start, and $k$ the length. Masking in the middle leads to stronger frequency-domain leakage via convolution with sinc functions and can destabilize training.

For completeness, we note the mid-segment masking variant as:
\begin{equation}
    	ilde{s}[n]=s[n]\cdot g_{n_0}[k]
    \label{eq:mask_from_middle}
\end{equation}
which in the frequency domain corresponds to a convolution with a sinc-like kernel:
\begin{equation}
    	ilde{S}(\omega)=S(\omega)\ast \mathrm{sinc}(\omega)
    \label{eq:mask_from_middle_freq}
\end{equation}

\begin{figure*}
    \centering
    \includegraphics[width=12cm]{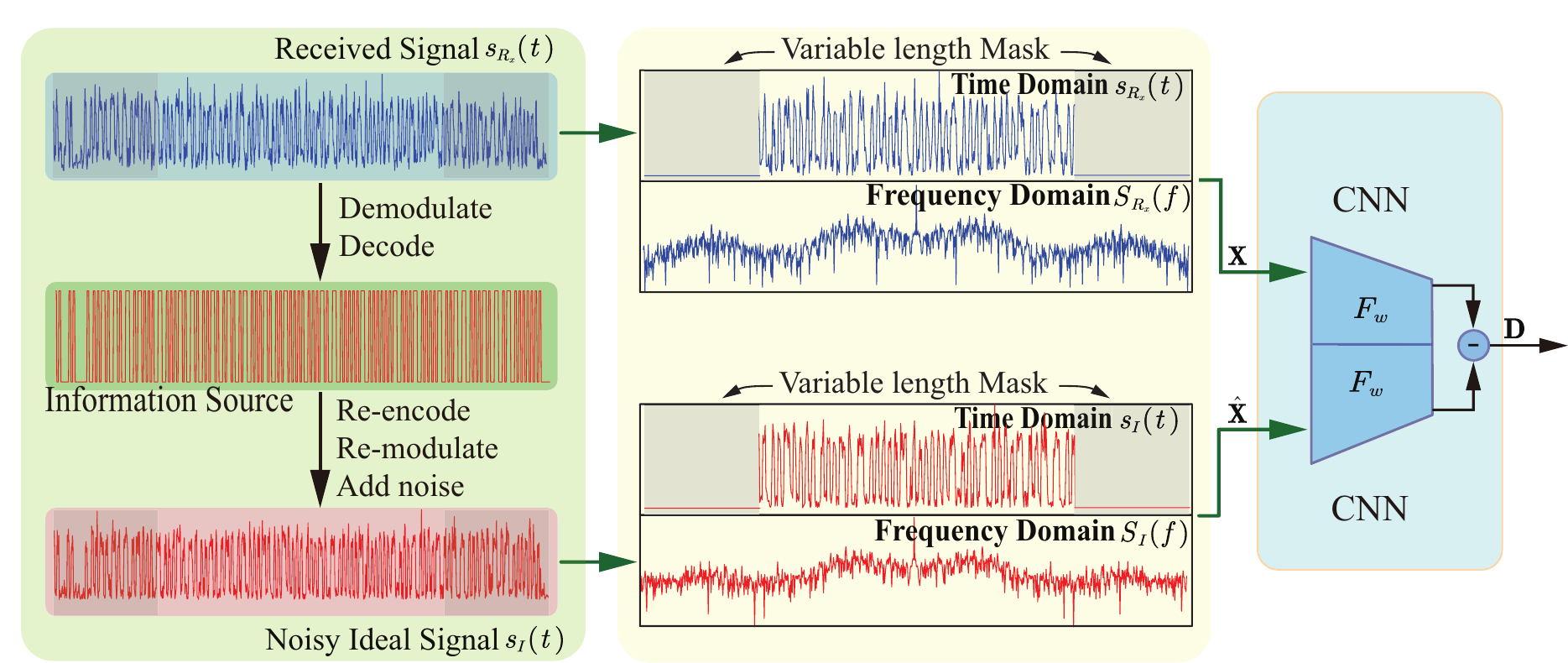}
    \caption{Data augmentation: random start/end time-domain masking}
    \label{fig:data_augmentation}
\end{figure*}

Our augmentation applies a variable-length mask at either the start or end of a sample (Fig.~\ref{fig:data_augmentation}). The mask length is drawn uniformly from 0 to one-sixth of the sample length and masked regions are zeroed.

The full exemplar-free incremental algorithm is given in Algorithm~\ref{alg:gmm_em}.
\begin{algorithm}[htbp]
    \caption{Exemplar-Free Class-Incremental Learning}
    \label{alg:gmm_em}
    \begin{algorithmic}[1]
        \REQUIRE Pretrained feature extractor $F_w$, training data $X,\hat{X}$, number of incremental rounds $T$
        \IF {initial training}
            \FOR{initial training}
            \STATE initialize $Adapter_0$ and $\phi^0$
            \STATE train classifier $\phi^0$ and $Adapter_0$ on $X,\hat{X}$ and extract per-class feature sets ${D_c^0}$
            \STATE fit diagonal GMMs per class and save parameters $\mathcal{N}_c$
            \STATE retain $Adapter_0$
            \ENDFOR
        \ELSE
            \FOR{incremental rounds $t=1$ to $T$}
            \STATE obtain new-class data $X_t,\hat{X}_t$
            \STATE initialize $Adapter_t$
            \STATE initialize classifier $\phi^t$ by copying previous classifier and apply weight alignment
            \STATE train $Adapter_t$ and classifier $\phi^t$ using new-class data $X_t,\hat{X}_t$
            \STATE fit GMM parameters for new classes and save $\mathcal{N}_t$
            \STATE initialize student Adapter $Adapter_s$ for distillation
            \STATE use all $Adapter_{0:t}$ as teachers; sample pseudo-features from historical GMMs $\mathcal{N}_{0:(t-1)}$ and use current data $X_t,\hat{X}_t$ to train $Adapter_s$ with feature alignment
            \STATE fine-tune classifier $\phi^t$ with sampled pseudo-features and current features
            \STATE use $Adapter_s$ and $\phi^t$ for inference
            \ENDFOR
        \ENDIF
    \end{algorithmic}
\end{algorithm}

\section{Experiments}
\subsection{Dataset}
The dataset comprises ADS-B signals captured with an AD9361 front-end at 1091 MHz (10 Msps sampling, 5 MHz bandwidth). We prepared two temporally disjoint collections: one for pretraining (Feb 19--26, 2025) and one for incremental evaluation (Jan 17--21, 2025). Saturated and low-SNR recordings were removed, and signals were segmented into fixed-length frames (112 samples after filtering). Only classes with more than 100 records were retained. The pretraining set contains 2,175 classes; the incremental set contains 669 classes. For experiments we used a 70/30 train/test split.
\begin{table}[!ht]
    \centering
    \begin{tabular}{l|l|l}
    \hline
        \textbf{Data} & \textbf{Number of classes} & \textbf{Train/Test split} \\
        \hline
        Pretraining & 2175 & 7:3  \\
        \hline
        Incremental & 669 & 7:3  \\
        \hline
    \end{tabular}
    \caption{Dataset statistics}
    \label{table:dataset}
\end{table}

\subsection{Experimental Setup}
\begin{figure}
    \centering
    \includegraphics[width=5cm]{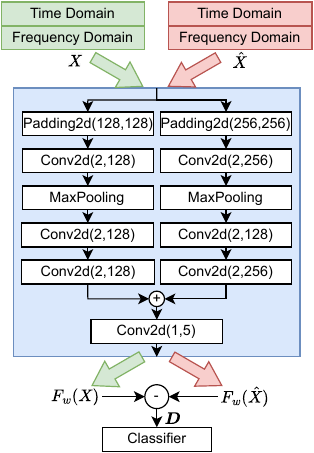}
    \caption{Structure of feature extractor}
    \label{fig:extractor_structure}
\end{figure}

All backbones are based on the twin-difference extractor proposed in \cite{jiangGeneralizedRadioFrequency2024}, pretrained on the pretraining split. The extractor architecture is shown in Fig.~\ref{fig:extractor_structure}. 
$F_w(\cdot)$ denotes the extractor mapping; $X$ and $\hat{X}$ correspond to original time-frequency representations and matched reconstructions without hardware fingerprints. The classifier is a fully-connected layer with 1D batch normalization as shown in Fig.~\ref{fig:classifier_structure}. For per-class diagonal GMM fitting, we set the number of components $k=2$.

\subsection{Results}
We first compare t-SNE visualizations of real and GMM-sampled pseudo-features. Five classes were randomly selected from the training set, and their features were computed and visualized using t-SNE (Fig.~\ref{subfig:tsne_orig}). Then, 500 pseudo-features were sampled from the GMMs and visualized (Fig.~\ref{subfig:tsne_pseudo}). Overlaying both visualizations (Fig.~\ref{subfig:tsne_overlay}) shows that GMM-generated pseudo-features largely cover the main geometric regions of real sample features, indicating that GMM sampling helps preserve decision boundaries.

\begin{figure}
    \centering
    \includegraphics[width=8cm]{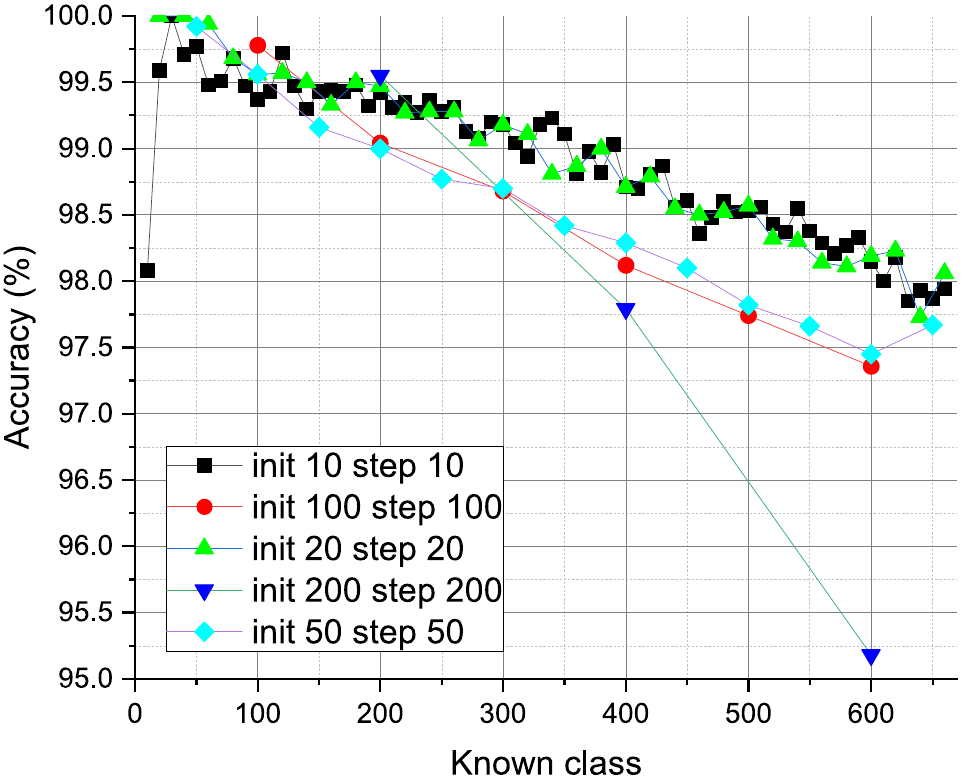}
    \caption{Incremental training results of the proposed method under different initial and incremental class counts}
    \label{fig:mycil_all}
\end{figure}

We use the PyCIL framework \cite{zhou2023pycil} and compare our method with EWC, Finetune, PASS \cite{zhuPrototypeAugmentationSelfSupervision2021}, iCaRL \cite{rebuffiICaRLIncrementalClassifier2017}, WA \cite{zhaoMaintainingDiscriminationFairness2019a}, Dynamically Expandable Representation \cite{yanDynamicallyExpandableRepresentation2021a}, and LwF. Fig.~\ref{fig:mycil_all} shows results under different initial and incremental class counts: the x-axis represents the number of classes learned, and the y-axis represents the test accuracy on all historical real samples (a measure of forgetting). Curves fluctuate but generally decline as incremental rounds progress. Scenarios with smaller initial and incremental steps perform better. After many incremental rounds, our method maintains high accuracy, demonstrating its effectiveness.

\begin{figure*}[h]
    \centering
    \subfloat[Start: 10 classes, Increment: 10 classes\label{subfig:10saving}]{\includegraphics[width=8cm]{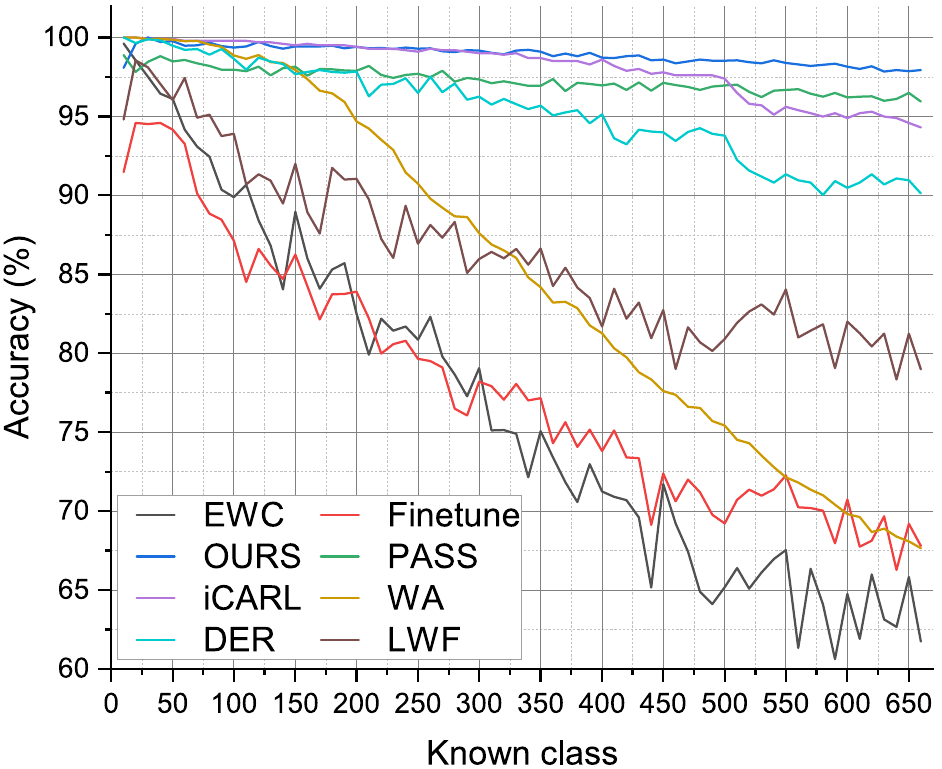}}
    \subfloat[Start: 20 classes, Increment: 20 classes\label{subfig:20saving}]{\includegraphics[width=8cm]{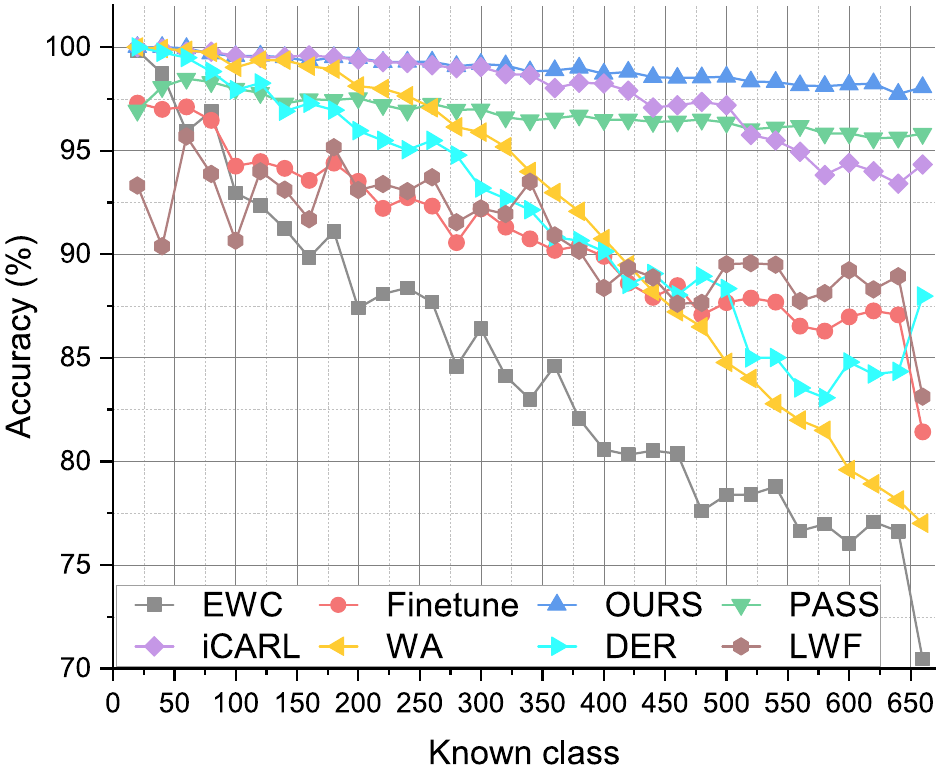}}
    \hfill
    \subfloat[Start: 50 classes, Increment: 50 classes\label{subfig:50saving}]{\includegraphics[width=8cm]{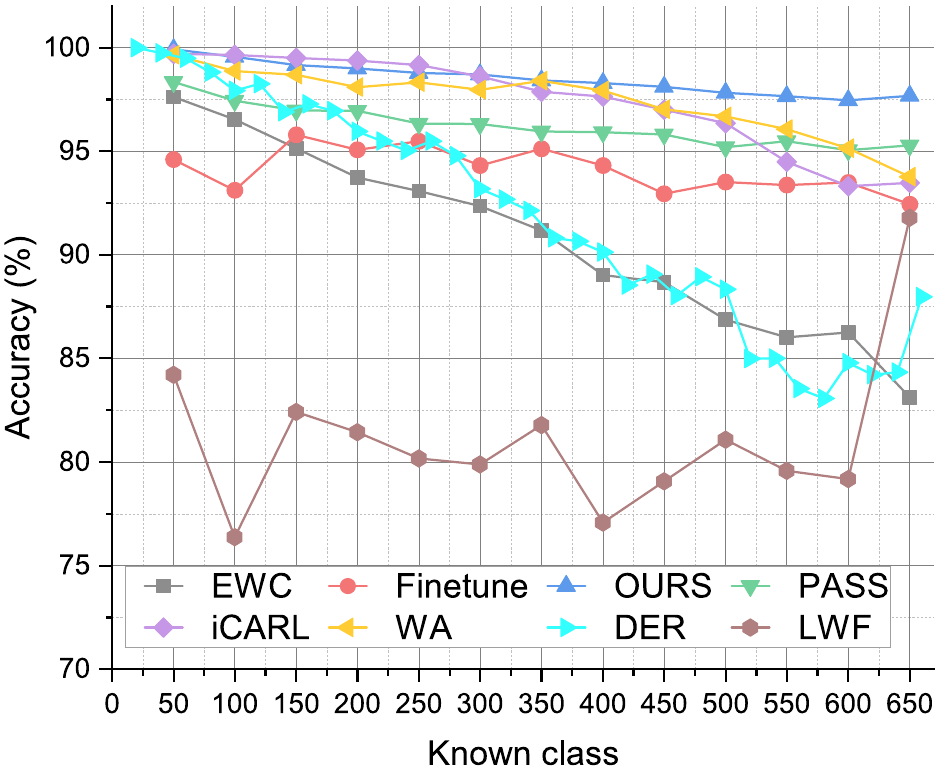}}
    \subfloat[Start: 100 classes, Increment: 100 classes\label{subfig:100saving}]{\includegraphics[width=8cm]{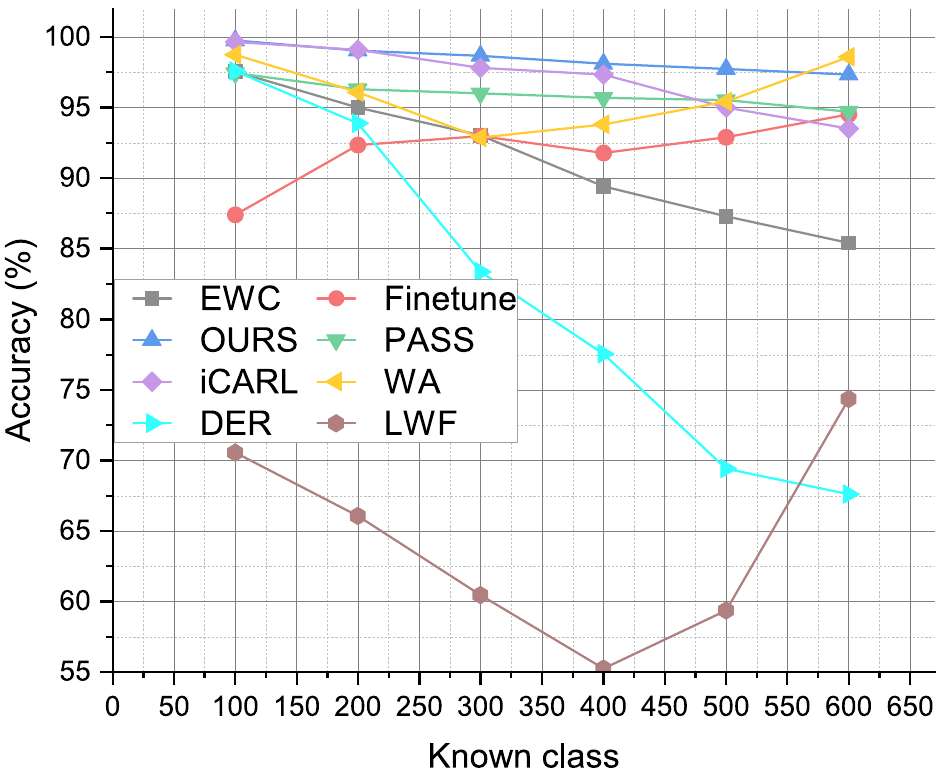}}
    \caption{Comparison of our method with other methods when exemplar samples are retained}
    \label{fig:cross_compare}
\end{figure*}

\begin{figure*}
    \centering
    \subfloat[Start: 10 classes, Increment: 10 classes\label{subfig:10nosaving}]{\includegraphics[width=8cm]{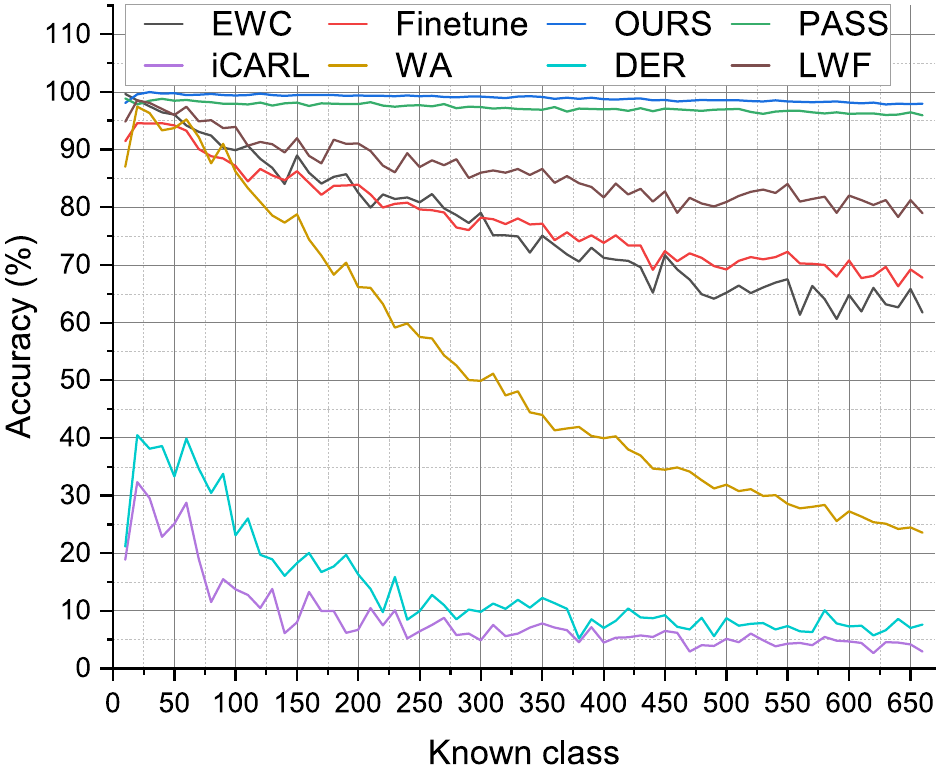}}
    \subfloat[Start: 20 classes, Increment: 20 classes\label{subfig:20nosaving}]{\includegraphics[width=8cm]{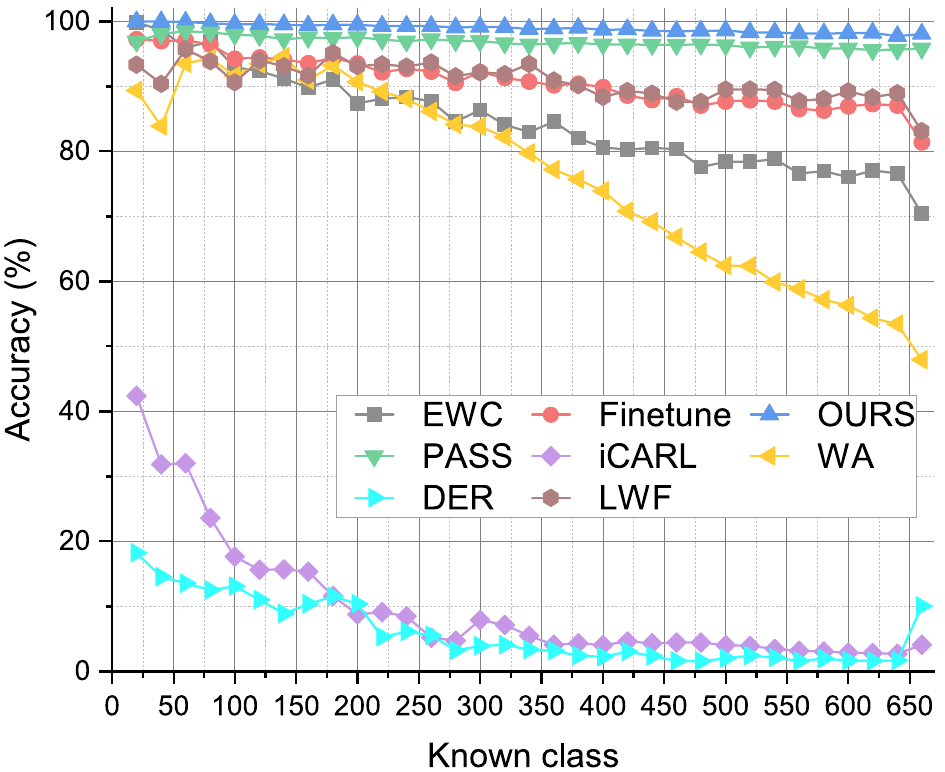}}
    \caption{Comparison of our method with other methods when no old exemplars are stored}
    \label{fig:cross_compare_nosaving}
\end{figure*}

Cross-method comparisons under exemplar-preserving and exemplar-free settings are shown in Fig.~\ref{fig:cross_compare} and Fig.~\ref{fig:cross_compare_nosaving}. In exemplar-keeping comparisons, all methods store 2000 exemplars when applicable; other methods do not store old samples. Our method consistently achieves the best results. When no exemplars are allowed (Fig.~\ref{fig:cross_compare_nosaving}), exemplar-dependent methods (DER, WA, iCaRL) degrade quickly.

\begin{figure}
    \centering
    \includegraphics[width=7cm]{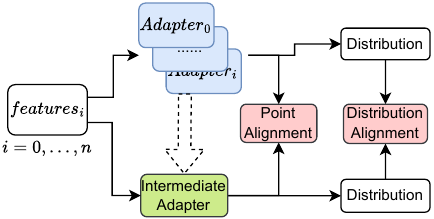}
    \caption{Distillation process merging previous Adapters into an intermediate Adapter during training}
    \label{fig:intermediate_alignment}
\end{figure}

We also compare extra storage overhead. In the B10S10 scenario, our method stores per-stage Adapters and per-class GMM parameters; storage per stage is approximately 25.52 KB (Table~\ref{tab:test2}). GMM storage per class is small (approximately 2.07 KB per class under the chosen settings), and Adapter size is about 4.82 KB per stage, which can be further reduced by distillation merging of Adapters across stages (Fig.~\ref{fig:intermediate_alignment}).

For reference, the per-class storage of diagonal-GMM parameters can be estimated as:
\begin{equation}
    \begin{aligned}
        Storage_{GMM} &= k \times (2 \times D + 1) \times bytes\;per\;float\\
        &= 2\times(2\times132+1)\times4 \\
        &= 2120 \,\,Bytes\approx 2.07 \,\,KB
    \end{aligned}
    \label{eq:gmm_storage}
\end{equation}
where $k$ is the number of mixture components, $D$ is the feature dimension, and $bytes\;per\;float$ is the storage per floating-point value.

\begin{table}
    \centering
    \begin{threeparttable}[b]
    \caption{Comparison of parameter/storage overhead across methods}
    \label{tab:test2}
    \begin{tabular}{ccccc}
        \hline
            Method   & Model size & Uses exemplars & Keeps old network & Extra storage           \\
            \hline
            $\mathbf{OURS}$     & 9385.7KB     & no            & no            & 25.52KB/stage\tnote{1} \\
            iCaRL    & 9438.4KB     & yes           & yes           & 0                           \\
            DER      & 9491.7KB     & yes           & no            & 2.3MB/stage\tnote{2}                       \\
            EWC      & 9438.4KB     & no            & no            & 9.66MB\tnote{3}                       \\
            Finetune & 9391.0KB     & no            & no            & 0                           \\
            WA       & 9396.2KB     & yes           & yes           & 0                           \\
            LwF      & 9391.0KB     & no            & yes           & 0                           \\
            PASS     & 9438.4KB     & no            & yes           & 9.2MB\tnote{4}                      \\
        \hline
    \end{tabular}
    \begin{tablenotes}
    \item[1] Store one Adapter per stage and one diagonal-GMM parameter set per class.
    \item[2] Expand a new feature extractor per stage.
    \item[3] Store one Fisher information matrix.
    \item[4] Store subspace bases and projection parameters per stage.
    \end{tablenotes}
    \end{threeparttable}
\end{table}

Table~\ref{table:accuracy_compare} compares final and mean accuracies across methods under different initial/incremental settings. Our method consistently outperforms baselines in average accuracy and long-term stability.

\begin{table*}[]
    \centering
    \caption{Accuracy comparison across methods}
    \label{table:accuracy_compare}
    \begin{tabular}{ccccccccc}
        \hline
        Method   & \multicolumn{2}{c}{B10S10} & \multicolumn{2}{c}{B20S20} & \multicolumn{2}{c}{B50S50} & \multicolumn{2}{c}{B100S100} \\ \hline
                & $A_l$   & $\bar{A}$   & $A_l$   & $\bar{A}$   & $A_l$   & $\bar{A}$   & $A_l$    & $\bar{A}$    \\ \hline
        $\mathbf{OURS}$     & 97.94            & 98.92   & 98.06            & 98.94   & 97.67            & 98.5    & 97.36             & 98.48    \\
        iCaRL    & 94.3             & 97.1    & 94.33            & 97.1    & 93.47            & 97.39   & 93.51             & 97.07    \\
        DER      & 90.12            & 95.31   & 87.97            & 91.89   & 82.56            & 86.27   & 67.59             & 81.56    \\
        EWC      & 61.74            & 76.6    & 70.45            & 84.66   & 83.13            & 90.74   & 85.4              & 91.3     \\
        Finetune & 67.79            & 77.97   & 81.42            & 90.77   & 92.45            & 90.77   & 94.53             & 92       \\
        WA       & 67.66            & 85.26   & 77               & 91.6    & 93.77            & 97.43   & 98.59             & 95.92    \\
        LwF      & 78.97            & 86.44   & 83.13            & 90.82   & 91.79            & 81.08   & 74.36             & 64.34    \\
        PASS     & 95.94            & 97.28   & 95.81            & 96.81   & 95.28            & 96.23   & 94.71             & 95.96    \\ \hline
    \end{tabular}
\end{table*}

\section{Conclusion}
This paper proposes an exemplar-free class-incremental learning framework for deep-learning-based RF fingerprint recognition. Leveraging a pretrained feature extractor, we fit per-class features with diagonal Gaussian Mixture Models (GMMs) and generate pseudo-features on demand to rehearse past classes during incremental training, eliminating the need to store raw samples. Extensive experiments on large ADS-B datasets demonstrate that our method significantly improves memory retention of old classes compared to mainstream baselines, validating its effectiveness and practicality for resource- and privacy-constrained scenarios.

\bibliographystyle{IEEEtran}
\bibliography{class_incremental}

\end{document}